\pdfoutput=1
\documentclass[fleqn,usenatbib,usedcolumn]{mnras}

\usepackage{graphicx,amssymb,hyperref}
\usepackage{multirow}
\usepackage{color}

\usepackage[fleqn]{amsmath}
\usepackage{abbrev}
\usepackage{siunitx}
\usepackage{comment} 
\usepackage{gensymb}
\usepackage{booktabs,tabularx}
\usepackage{flushend}
\usepackage{xspace}
\usepackage{physics}
 \usepackage{xcolor}
 \usepackage{soul}
 
\usepackage{nicefrac}
\usepackage{units}
\usepackage{verbatim}
\usepackage[shortlabels]{enumitem}
\usepackage{subcaption}
\setlist[itemize]{leftmargin=5pt}

\def\mnras{MNRAS}
\def\apj{ApJ}

\def\simlt{\lower.5ex\hbox{$\; \buildrel < \over \sim \;$}}
\def\simgt{\lower.5ex\hbox{$\; \buildrel > \over \sim \;$}}

\def\kpc{\mathrm{\, kpc}}

\def\mpc{\mathrm{\, Mpc}}
\def\gpc{\mathrm{\, Gpc}}
\def\msun{\mathrm{\, M_\odot}}

\defcitealias{2012ApJ...761....1W}{W12}
\defcitealias{2019MNRAS.488.3646R}{R19}
\defcitealias{2020Sci...369.1347M}{M20}
\newcommand{\Men}{\citetalias{2020Sci...369.1347M}\xspace}

\newcommand{\eagle}{\textsc{eagle}\xspace}
\newcommand{\ceagle}{\textsc{c-eagle}\xspace}
\newcommand{\pysph}{\textsc{pysphviewer}\xspace}

\def\gs{\mathrel{\raise1.16pt\hbox{$>$}\kern-7.0pt \lower3.06pt\hbox{{$\scriptstyle \sim$}}}}         
\def\ls{\mathrel{\raise1.16pt\hbox{$<$}\kern-7.0pt \lower3.06pt\hbox{{$\scriptstyle \sim$}}}}   

\newcommand{\vect}[1]{\boldsymbol{#1}}
\newcommand{\code}[1]{\texttt{#1}}

\newcommand{\be}{\begin{equation}}
\newcommand{\ee}{\end{equation}}
\newcommand{\ba}{\begin{eqnarray}}
\newcommand{\ea}{\end{eqnarray}}

\DeclareGraphicsExtensions{.pdf,.png}

\title[GGSL of simulated clusters]{The galaxy-galaxy strong lensing cross-sections of simulated $\Lambda$CDM galaxy clusters}

\author[A.\ Robertson]{Andrew Robertson\thanks{e-mail: {\tt andrew.robertson@durham.ac.uk}}\\Institute for Computational Cosmology, Durham University, South Road, Durham DH1 3LE, UK}

\begin{document}

\maketitle

\label{firstpage}

\begin{abstract}

We investigate a recent claim that observed galaxy clusters produce an order of magnitude more galaxy-galaxy strong lensing (GGSL) than simulated clusters in a $\Lambda$CDM cosmology. We take galaxy clusters from the \ceagle hydrodynamical simulations and calculate the expected amount of GGSL for sources placed behind the clusters at different redshifts. The probability of a source lensed by one of the most massive \ceagle clusters being multiply imaged by an individual cluster member is in good agreement with that inferred for observed clusters. We show that numerically converged results for the GGSL probability require higher resolution simulations than had been used previously. On top of this, different galaxy formation models predict cluster substructures with different central densities, such that the GGSL probabilities in $\Lambda$CDM cannot yet be robustly predicted. Overall, we find that galaxy-galaxy strong lensing within clusters is not currently in tension with the $\Lambda$CDM cosmological model.

\end{abstract}


\section{Introduction}

Structure within a $\Lambda$ Cold Dark Matter (CDM) universe builds up hierarchically, with massive clusters of galaxies formed from mergers of lower-mass clusters and groups, as well as individual galaxies. This merging process is incomplete, such that galaxy clusters contain a large amount of self-bound substructure, known as subhaloes. The number of these subhaloes can be probed using strong gravitational lensing, which can then be compared with predictions for a given cosmological model \citep[e.g.][]{2017MNRAS.468.1962N}. Additional information can be extracted from looking at the internal structure of subhaloes, which in a CDM-only universe can be robustly predicted using $N$-body simulations \citep[e.g.][]{2012MNRAS.425.2169G}. However, the inclusion of baryons into simulations modifies these predictions.


\citet[][hereafter  \citetalias{2020Sci...369.1347M}]{2020Sci...369.1347M} compared the subhalo properties of observed galaxy clusters (inferred from strong gravitational lensing) with those of a mass-matched sample of clusters from cosmological hydrodynamical simulations \citep{2014MNRAS.438..195P}. They found that subhaloes within observed clusters appeared to be more centrally concentrated than those in simulated clusters. This was summarised as a factor of approximately ten discrepancy in the expected amount of galaxy-galaxy strong lensing (GGSL) within clusters, where GGSL refers to background sources being multiply imaged due to gravitational lensing, with the image splitting being done by an individual galaxy within the cluster as opposed to by the cluster as whole. \citetalias{2020Sci...369.1347M} concluded that either dark matter was something other than CDM, or that there are systematic issues with the way cosmological simulations are currently performed.

In this paper we use a different set of hydrodynamical simulations from the simulations used in \citetalias{2020Sci...369.1347M}  to further assess the compatibility of $\Lambda$CDM with observed clusters. In Section~\ref{sect:sims} we discuss the hydrodynamical simulations and the method we employ to measure their strong lensing properties. We then compare the simulated clusters with observed clusters in Section~\ref{sect:results}, before concluding in Section~\ref{sect:conclusions}. Our simulations and lensing analysis all assume the cosmology from \citet{2014A&A...571A..16P}.

\section{Gravitational lensing from simulated galaxy clusters}
\label{sect:sims}


The \ceagle project \citep{2017MNRAS.470.4186B,2017MNRAS.471.1088B} uses a zoom simulation technique to resimulate (at higher resolution) galaxy cluster haloes found in a parent simulation with a side length of $3.2 \gpc$ \citep{2017MNRAS.465..213B}. The high-resolution region around each cluster is selected so that no lower resolution particles are present within a radius of $5 \, r_{200}$ from the cluster centre at $z = 0$, with a subset of the \ceagle haloes having high-resolution regions that extend out to $10 \, r_{200}$ \citep[this subset is known as `Hydrangea';][]{2017MNRAS.470.4186B}. The high-resolution region around each cluster matches the resolution of the \eagle$100 \mpc$ simulation \citep[Ref-L100N1504,][]{2015MNRAS.446..521S}, with DM particle mass $m_\mathrm{DM} = \num{9.7e6} \msun$ and initial gas particle mass $m_\mathrm{gas} = \num{1.8e6} \msun$. They use the \eagle galaxy formation model \citep{2015MNRAS.446..521S,2015MNRAS.450.1937C}, which includes radiative cooling, star formation, stellar evolution, feedback due to stellar winds and supernovae, and the seeding, growth and feedback from black holes. The specific calibration of \eagle that was used is labelled as `AGNdT9' in \citet{2015MNRAS.446..521S}. 

We analyse the full sample of 30 \ceagle clusters, which at $z=0$ have a uniform distribution of $\log_{10} (M_{200} / \msun)$ spanning from 14 to 15.4. We use the $z=0$ snapshots because these are available for all of the \ceagle clusters, but place the clusters at a lens redshift of $z_\mathrm{l}=0.4$ as this was the median redshift of the observed clusters in \citetalias{2020Sci...369.1347M}. We found that using the $z=0$ snapshots does not substantially bias our results compared with using clusters extracted at the true lens redshift, which we tested using the 16 clusters that had a $z=0.411$ snapshot.\footnote{Specifically, the $z_\mathrm{snap}=0.411$ GGSL cross-sections divided by those with $z_\mathrm{snap}=0$ have an approximately log-normal distribution, with $\langle \log_{10} \sigma_\mathrm{GGSL}^{z=0.411} / \sigma_\mathrm{GGSL}^{z=0.0} \rangle=-0.01$ and $\langle \left(\log_{10} \sigma_\mathrm{GGSL}^{z=0.411} / \sigma_\mathrm{GGSL}^{z=0.0}\right)^2 \rangle=0.21$, corresponding to little bias and a scatter of about $0.4 \, \mathrm{dex}$.}

The observed sample with which we will later compare (`Ref.' from \citetalias{2020Sci...369.1347M}) is comprised of three clusters at redshifts of 0.44, 0.40 and 0.35, with weak-lensing inferred halo masses of $M_{200} = 1.6, 1.0$ and $2.0 \times 10^{15} \msun$, respectively \citep{2014ApJ...795..163U}. For comparison, there are seven \ceagle clusters with $M_{200} > 10^{15} \msun$, with the most massive halo being $\num{2.4e15} \msun$. This means that the most massive \ceagle clusters are comparable (at least in terms of their halo masses) with the observed sample.


\subsection{Calculating deflection angles}
\label{sect:deflection_angles}

In order to calculate the GGSL cross-sections of our simulated galaxy clusters we need to first calculate their deflection angle fields. To begin, we calculate projected density maps of each particle type (DM, gas, stars and black holes) independently. We take all particles identified as part of the main friends-of-friends group within the zoom-in region and calculate a $4 \times 4 \mpc^2$ projected density field using \pysph \citep{alejandro_benitez_llambay_2015_21703}, projecting along the simulation $z$-axis. The maps have 4096 pixels on a side, corresponding to a pixel scale of roughly $1 \kpc$. \pysph smooths the mass of each particle using a smoothing kernel, with the kernel size for each particle set as the distance to the $n_\mathrm{ngb}$th nearest neighbour (of the same particle type). Following \citetalias{2020Sci...369.1347M} we use $n_\mathrm{ngb}=50$, but we also impose a maximum smoothing length of $100 \kpc$ to speed up the calculation in low density regions (which are unimportant for the GGSL cross-section). The projected density maps for each particle type are summed up to produce the total projected density map for each cluster, $\Sigma$.

\begin{figure*}
        \centering
        \includegraphics[width=0.49\textwidth]{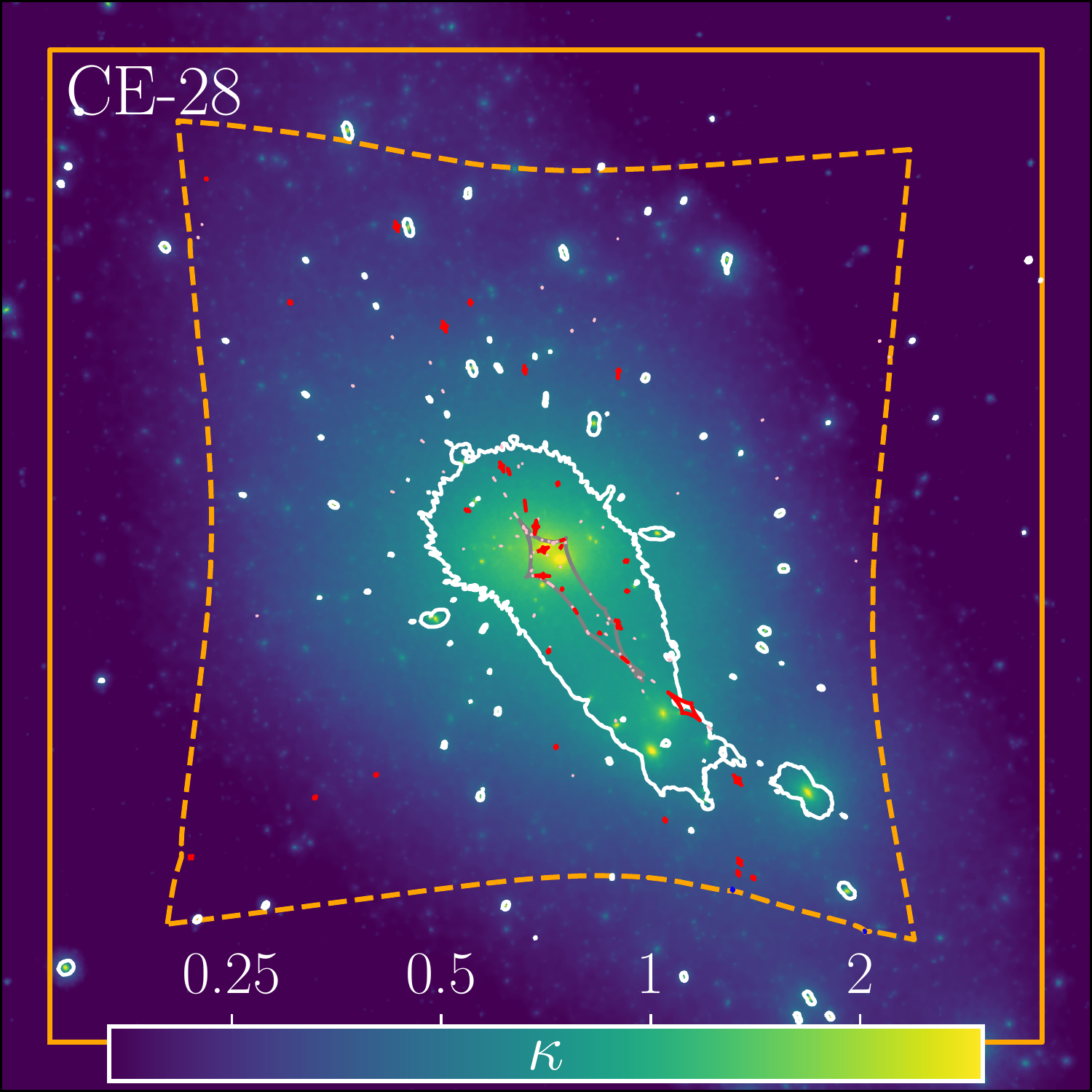}   
         \includegraphics[width=0.49\textwidth]{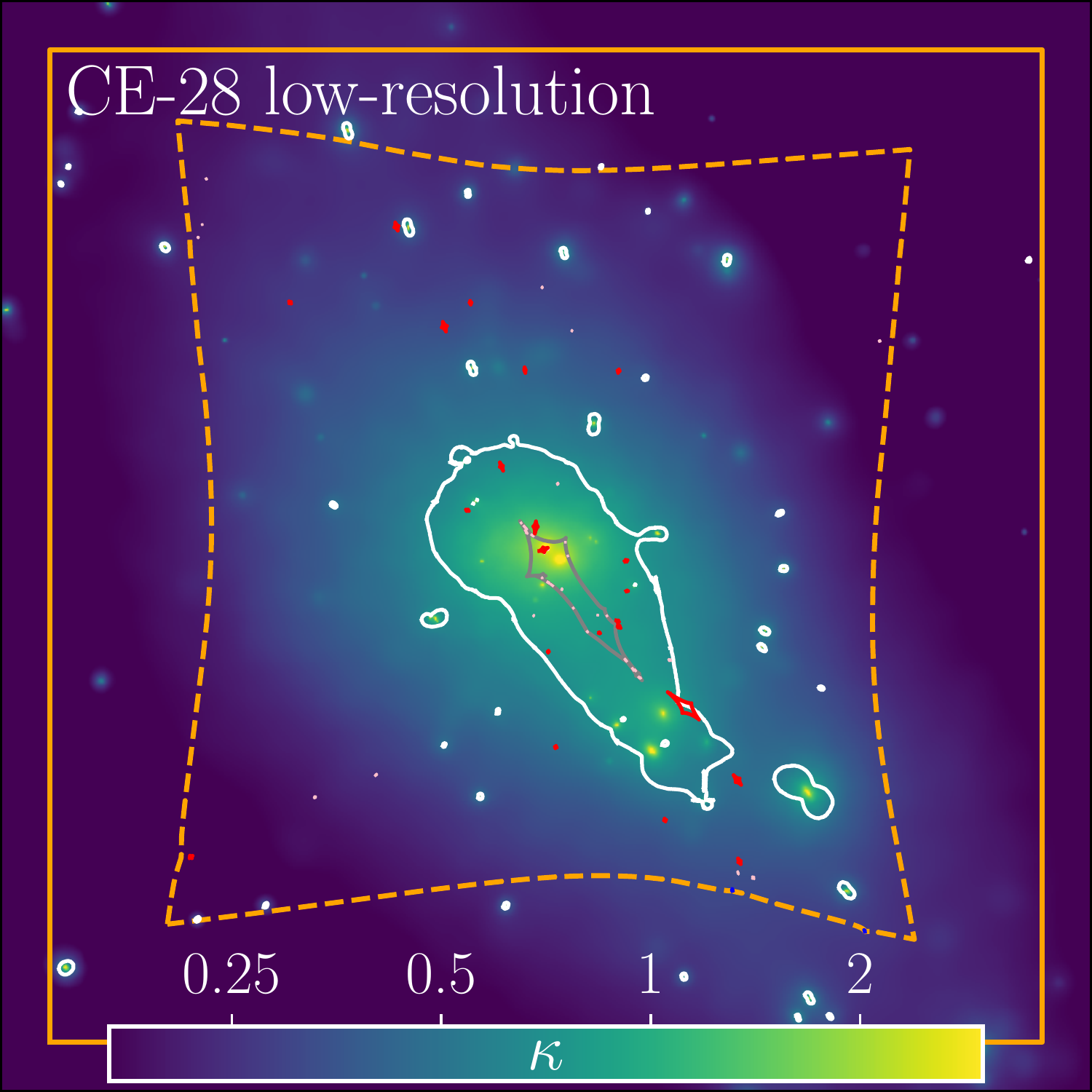}       
	\caption{The background image shows the convergence of one of our simulated clusters (CE-28), assuming lens and source redshifts of 0.4 and 7 respectively. The field-of-view that is plotted is 220 arcsec, with the solid orange box corresponding to the $200 \, \mathrm{arcsec}$ square region in which we calculate strong lensing quantities. Tangential critical curves are shown as solid white lines. Cluster-scale caustics are plotted as grey lines, and secondary caustics with $\theta_\mathrm{E,eff} < 0.5 \, \rm{arcsec}$ are plotted in pink. The secondary caustics that contribute to the GGSL cross-section are plotted in red. The dashed orange line corresponds to the solid orange line when mapped into the source plane, and it is within this source-plane region that we calculate the GGSL probability. The left panel is for the full-resolution simulation of CE-28, while the right panel corresponds to a simulation with 128 times worse mass resolution (similar to the \citet{2014MNRAS.438..195P} simulations used by \citetalias{2020Sci...369.1347M}), created from subsampling particles from the full simulation. Small density peaks that lead to GGSL in the left-panel are often smoothed over in the right-panel, showing the importance of high-resolution simulations for calculating the expected GGSL cross-section.}
	\label{fig:convergence}
\end{figure*}

The redshifts of the lens and of a background source define the critical surface density for lensing,
\begin{equation}
\Sigma_\mathrm{crit} = \frac{c^2}{4 \pi G} \frac{D_\mathrm{s}}{D_\mathrm{l}D_\mathrm{ls}},
\label{sigma_crit}
\end{equation}
where $D_\mathrm{s}$, $D_\mathrm{l}$, and $D_\mathrm{ls}$ are the angular diameter distances between the observer and source, observer and lens, and lens and source respectively. It is useful for lensing to define the dimensionless convergence, $\kappa= \Sigma / \Sigma_\mathrm{crit}$. An example convergence map is shown in Fig.~\ref{fig:convergence}. Both $\kappa$ and the deflection angle field, $\vect{\alpha}$, depend on spatial derivatives of the projected gravitational potential. This means that the relationship between the Fourier transforms of $\kappa$ and $\vect{\alpha}$ is a simple one, and we calculate $\vect{\alpha}$ from $\kappa$ using discrete Fourier transforms \citep[this is described in more detail in][]{2019MNRAS.488.3646R}. To mitigate the effects of periodicity that are inherent to using discrete Fourier transforms we zero-pad the $\kappa$ field by a factor of four in each direction (i.e. out to $16 \mpc$ on a side).

Having calculated $\vect{\alpha}$ on the same square grid as we originally calculated the projected density, we now increase the resolution and decrease the field-of-view of the deflection angle map. We do this using bicubic interpolation,\footnote{We used \code{scipy.interpolate.RectBivariateSpline}} making an $8192 \times 8192$ grid of deflection angles covering the central $200 \times 200 \, \rm{arcsec}^2$ region, which is the field-of-view used by \Men. $200 \, \rm{arcsec}$ corresponds to approximately $1.1 \mpc$  at our adopted lens redshift.

\subsection{Critical curves and caustics}
\label{sect:caustics}

Labelling the $x$ and $y$ components of $\vect{\alpha}$ as $\alpha_x$ and $\alpha_y$, respectively, the two components of the gravitational shear are
\begin{equation}
\gamma_1= \frac{1}{2} \left( \pdv{\alpha_x}{x} - \pdv{\alpha_y}{y}\right), \,\, \gamma_2 = \pdv{\alpha_x}{y} = \pdv{\alpha_y}{x}
\label{eq:gamma1}
\end{equation}
with the magnitude of the shear $\gamma = \sqrt{\gamma_1^2 + \gamma_2^2}$. The magnification is given by
\begin{equation}
\mu =\frac{1}{(1 - \kappa)^2 - \gamma^2} = \frac{1}{(1 - \kappa - \gamma)(1- \kappa + \gamma)}.
\label{eq:mu}
\end{equation}
Critical curves are regions of the image plane in which the magnification is infinite. From equation~\eqref{eq:mu} it can be seen that this happens when $\lambda_\mathrm{t} \equiv 1 - \kappa - \gamma = 0$ or $\lambda_\mathrm{r} \equiv 1- \kappa + \gamma = 0$. The first of these two cases produces \emph{tangential} critical curves, with images formed close to tangential critical curves strongly distorted tangentially to the critical curve. 
Following \citetalias{2020Sci...369.1347M}, it is the tangential critical curves that we are interested in for the GGSL cross-section, so we find the tangential critical curves using the \emph{marhcing squares} algorithm\footnote{We use \code{find\_contours} in scikit-image \citep{scikit-image}.} 
on a map of $\lambda_\mathrm{t}$. In Fig.~\ref{fig:convergence} these are plotted as white lines.

For the purposes of GGSL within clusters, we need to distinguish between cluster-scale critical curves and secondary critical curves. For each cluster we define the critical curve enclosing the largest area, as well as any critical curves with an effective Einstein radius\footnote{$\theta_\mathrm{E,eff}$ is the radius of a circle enclosing the same area as that enclosed by the critical curve \citep{2013SSRv..177...31M}.} $\theta_\mathrm{E,eff} > 5 \, \rm{arcsec}$, as primary critical curves associated with cluster-scale lensing. All others critical curves are deemed to be secondary critical curves associated with GGSL.

Critical curves when mapped into the source plane by the lens equation are known as caustics. Primary caustics correspond to cluster-scale critical curves mapped into the source plane, and are plotted as grey lines in Fig.~\ref{fig:convergence} (in this example there is only one). The secondary caustics are the result of mapping all other tangential critical curves into the source plane, they are plotted as red and pink lines.

\subsection{The GGSL cross-section and probability}

The GGSL cross-section is defined as the area within the source plane in which a galaxy would be multiply imaged by an individual cluster member. Approximating source galaxies as point sources, this can be calculated from the area enclosed within the secondary caustics, as done in \citetalias{2020Sci...369.1347M}.  Following \Men we exclude caustics corresponding to secondary critical curves with $\theta_\mathrm{E,eff} < 0.5 \, \rm{arcsec}$ from the GGSL cross-section, because these produce GGSL events that would be hard to identify observationally.

The GGSL probability is calculated by dividing the GGSL cross-section by some nominal total source plane area. In \citetalias{2020Sci...369.1347M} this total area was the area of a $200 \times 200 \, \mathrm{arcsec}^2$  square in the image plane, mapped into the source plane, and we use the same definition here. 

\subsection{Changing the simulation resolution}
\label{sect:subsample}

In order to investigate how the GGSL cross-section and probability are impacted by simulation resolution, we used a subsampling technique to generate lower-resolution versions of the \ceagle clusters. With a subsample factor, $S$, we select only one in every $S$ particles from the original simulation, and multiply the selected particles' masses by a factor of $S$. This produces a realisation of the cluster with a mass resolution that is worse than the original by a factor of $S$.

Starting with this subsampling procedure, we can then generate lensing maps and calculate the corresponding GGSL cross-sections following the same procedures as for the full resolution simulation. An example is shown in the right panel of Fig.~\ref{fig:convergence}, where we show the same cluster as in the left panel, but subsampled with $S=128$. It is clear that as we reduce the mass resolution, the mass maps and critical curves become smoother, because they are made by smoothing particles with a kernel size equal to the distance to the 50th nearest neighbour. With fewer particles, the distances between particles increase, and so the smoothing lengths increase. 

Note that a property converging with respect to varying $S$ is necessary but not sufficient to show that the property is converged with respect to the original resolution of the simulations. This is because varying $S$ only checks for convergence of the lensing calculations with changing resolution, whereas the simulated mass distributions themselves could systematically change with resolution (beyond just changes to the amount of particle noise). 

\section{Results}
\label{sect:results}

\begin{figure}
        \centering
        \includegraphics[width=\columnwidth]{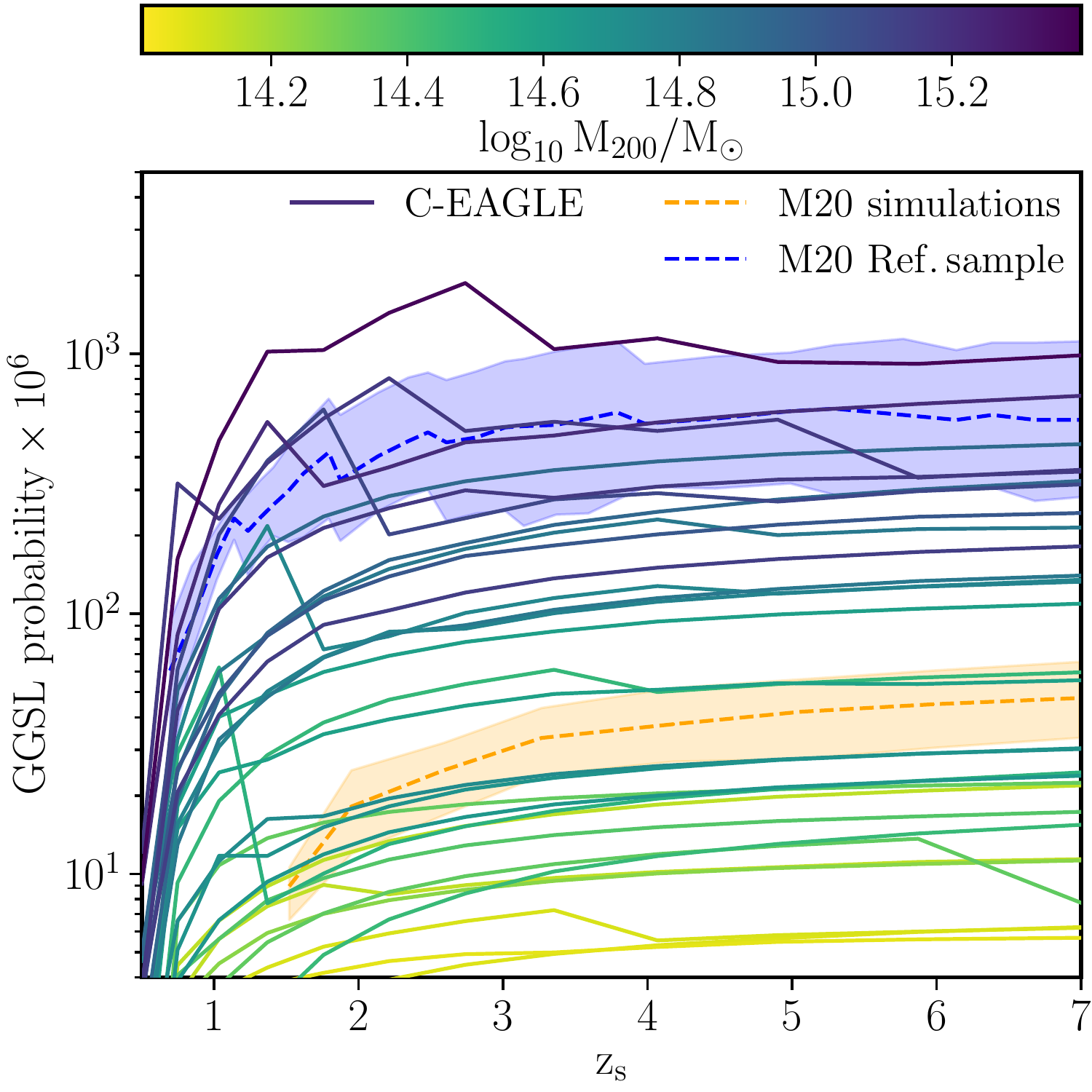}   
	\caption{The GGSL probability as a function of source redshift. The dashed lines and associated shaded regions are from \citetalias{2020Sci...369.1347M}, while the solid lines are for the \ceagle clusters, with the colour indicating the halo mass (see the colourbar at the top).}
	\label{fig:GGSLprob}
\end{figure}


In Fig.~\ref{fig:GGSLprob} we plot the GGSL cross-section as a function of source redshift for the 30 \ceagle clusters, as well as for the simulated and observed samples from \citetalias{2020Sci...369.1347M}. Note that we only show the `Ref' observational sample for clarity, but the other two observed samples had similar GGSL probabilities. The \ceagle GGSL probabilities vary by over two orders of magnitude, with an unsurprising trend that more massive clusters tend to produce more GGSL. Even haloes with similar masses show a non-negligible spread in GGSL probability, such that a detailed comparison between the observations and simulations would need to consider cluster properties beyond simply halo mass. The most massive \ceagle clusters have similar GGSL probabilities to the observed clusters. \emph{There is no obvious discrepancy between observed galaxy clusters and those simulated in the context of $\Lambda$CDM.}

\subsection{Dependence of the GGSL probability on simulation resolution}

\begin{figure*}
        \centering
        \includegraphics[width=\textwidth]{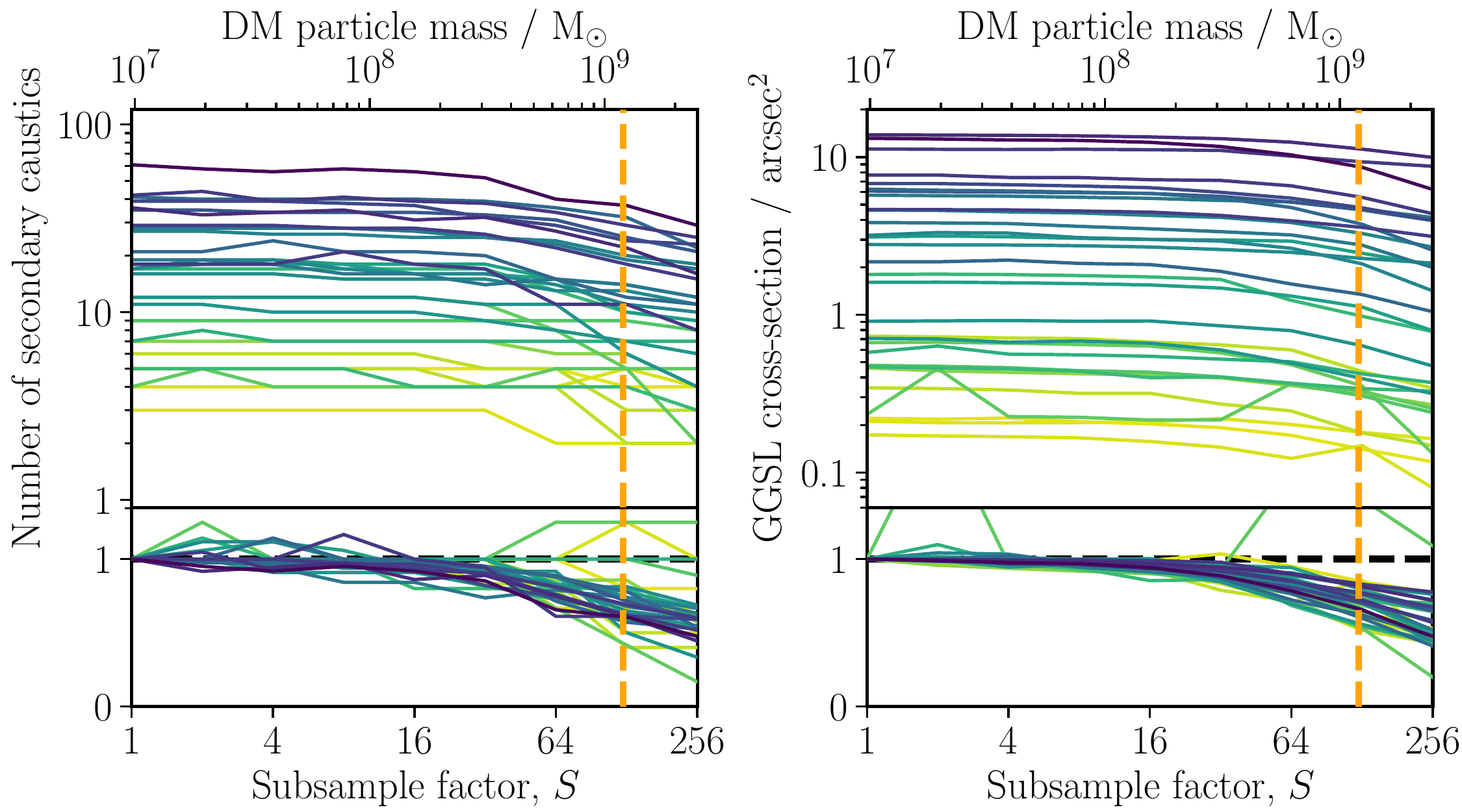}   
	\caption{The number of secondary caustics with $\theta_\mathrm{E,eff} > 0.5 \, \rm{arcsec}$ (left panel) and the area enclosed within those secondary caustics (the `GGSL cross-section', right panel) as a function of the effective resolution of our simulations. The lens and source redshifts are $z_\mathrm{l} = 0.4$ and $z_\mathrm{s} = 7$, respectively. The results for all 30 \ceagle clusters are plotted, with darker colours corresponding to more massive haloes (using the colour scheme from Fig.~\ref{fig:GGSLprob}). The bottom panels show the values relative to the highest-resolution ($S=1$) value for each cluster. At the top of the plot the resolution is expressed in terms of the DM particle mass. \citetalias{2020Sci...369.1347M}'s fiducial simulations used a DM particle mass of $\num{1.2e9} \msun$, denoted by the dashed orange lines. Note that the odd behaviour of the GGSL cross-section with $S$ for one of the low-mass clusters occurs because a large secondary critical curve separates from (and later merges with) the primary critical curve as $S$ is increased.
	}
	\label{fig:resolution}
\end{figure*}

The \ceagle simulations have around two orders of magnitude better mass resolution than the simulations in \Men. Using the subsampling procedure defined in Section~\ref{sect:subsample} we can asses whether the lower GGSL probabilities in the \Men simulations compared with \ceagle are driven by resolution. In Fig.~\ref{fig:resolution} we plot both the number of secondary caustics, and the total area enclosed within the secondary caustics (the GGSL cross-section), as a function of subsampling factor, $S$. The lower panels show these same quantities relative to their values with $S=1$, giving an indication of how we expect these quantities to vary with simulation resolution.

Both the number of secondary caustics and the GGSL cross-sections appear to be numerically converged at \ceagle resolution. At the resolution of the \Men simulations, neither the GGSL cross-sections nor the number of secondary caustics are converged, and from the \ceagle results we would expect a $20-50 \%$ reduction in the GGSL cross-section due to the \Men resolution, with a similar reduction in the number of secondary caustics. This goes some way to explaining the low GGSL cross-section of the \Men simulations, although it is clearly not the entire cause given the order of magnitude discrepancy between the \Men simulations and the largest GGSL cross-section \ceagle clusters.

\subsection{Subhalo concentrations}

The remaining discrepancy between the GGSL probabilities for \ceagle and the \Men simulations is explained by the fact that the \ceagle subhaloes are more concentrated than those in the \Men simulations \citep{BaheTemp}. This can be expressed in terms of \ceagle subhaloes having a higher maximum circular velocity, $v_\mathrm{max}$, at a given subhalo mass. For a singular isothermal sphere with a small core (which is approximately the density profile with which \Men model their subhaloes) the strong-lensing cross-section is proportional to $v_\mathrm{max}^4$ \citep{1987ApJ...320..468H}. As such, even modest changes to the central concentration of subhaloes will have substantial effects on the probability of GGSL.


\section{Conclusions}
\label{sect:conclusions}

Our primary finding is that state of the art simulations of galaxy clusters in a $\Lambda$CDM universe can produce clusters with GGSL probabilities in good agreement with those inferred for observed clusters. This agreement is in stark contrast to the order of magnitude discrepancy shown by \Men between the same observations and the simulations of \citet{2014MNRAS.438..195P}. The resolution of the \citet{2014MNRAS.438..195P} simulations is not sufficient to accurately predict the GGSL probability. Even with higher resolution, different hydrodynamical simulations currently make different predictions for the internal structure of subhaloes  \citep{BaheTemp}. 


It is not yet possible to use the GGSL probability to constrain the nature of dark matter. This is not to say that $\Lambda$CDM is correct, and instead indicates that more work is required before hydrodynamical simulations can make robust predictions for the internal structures of subhaloes within galaxy clusters. Going forward, the probability of GGSL within clusters may well prove a useful test of the $\Lambda$CDM model, but care will have to be taken to ensure that the theoretical predictions are converged with respect to the numerical resolution of simulations used, and that uncertainties in baryonic physics are taken into account. It would also be instructive to test the methods used to infer the GGSL probabilities of observed clusters on mock lensing data generated from simulations, in which the truth is known.

\section*{Acknowledgments}

Thanks to the \ceagle team for allowing me to use the \ceagle simulation data, and Alejandro Ben\'itez-Llambay for help with using \pysph. Thanks also to Yannick Bah\'e and Richard Massey, as well as Massimo Meneghetti, Priyamvada Natarajan, and the \Men team, for helpful comments on an earlier version of this manuscript. AR is supported by the European Research Council's Horizon2020 project `EWC' (award AMD-776247-6). This work used the DiRAC@Durham facility. The equipment was funded by STFC grants ST/K00042X/1, ST/P002293/1, ST/R002371/1, ST/S002502/1 and ST/R000832/1.

This work made use of the following software: \texttt{Astropy} \citep{2013A&A...558A..33A}, \texttt{IPython} \citep{2007CSE.....9c..21P}, \texttt{Matplotlib} \citep{2007CSE.....9...90H}, \texttt{NumPy} \citep{2011CSE....13b..22V}, \texttt{Py-SPHViewer} \citep{alejandro_benitez_llambay_2015_21703} and \texttt{SciPy} \citep{2011CSE....13b..22V}. 

\section*{Data availability}
Requests for the \ceagle data should be made to Yannick Bah\'e and David Barnes. The mass maps, deflection angle fields and corresponding critical curves and caustics will be shared on reasonable request to the author.

\bibliographystyle{mnras}

\bibliography{bibliography}

\begin{thebibliography}{}
\makeatletter
\relax
\def\mn@urlcharsother{\let\do\@makeother \do\$\do\&\do\#\do\^\do\_\do\%\do\~}
\def\mn@doi{\begingroup\mn@urlcharsother \@ifnextchar [ {\mn@doi@}
  {\mn@doi@[]}}
\def\mn@doi@[#1]#2{\def\@tempa{#1}\ifx\@tempa\@empty \href
  {http://dx.doi.org/#2} {doi:#2}\else \href {http://dx.doi.org/#2} {#1}\fi
  \endgroup}
\def\mn@eprint#1#2{\mn@eprint@#1:#2::\@nil}
\def\mn@eprint@arXiv#1{\href {http://arxiv.org/abs/#1} {{\tt arXiv:#1}}}
\def\mn@eprint@dblp#1{\href {http://dblp.uni-trier.de/rec/bibtex/#1.xml}
  {dblp:#1}}
\def\mn@eprint@#1:#2:#3:#4\@nil{\def\@tempa {#1}\def\@tempb {#2}\def\@tempc
  {#3}\ifx \@tempc \@empty \let \@tempc \@tempb \let \@tempb \@tempa \fi \ifx
  \@tempb \@empty \def\@tempb {arXiv}\fi \@ifundefined
  {mn@eprint@\@tempb}{\@tempb:\@tempc}{\expandafter \expandafter \csname
  mn@eprint@\@tempb\endcsname \expandafter{\@tempc}}}

\bibitem[\protect\citeauthoryear{{Astropy Collaboration} et~al.,}{{Astropy
  Collaboration} et~al.}{2013}]{2013A&A...558A..33A}
{Astropy Collaboration} et~al., 2013, \mn@doi [\aap]
  {10.1051/0004-6361/201322068}, \href
  {https://ui.adsabs.harvard.edu/abs/2013A&A...558A..33A} {558, A33}

\bibitem[\protect\citeauthoryear{{Bah{\'e}}}{{Bah{\'e}}}{2021}]{BaheTemp}
{Bah{\'e}} Y.~M.,  2021, arXiv e-prints, \href
  {https://ui.adsabs.harvard.edu/abs/2021arXiv210112112B} {p. arXiv:2101.12112}

\bibitem[\protect\citeauthoryear{{Bah{\'e}} et~al.,}{{Bah{\'e}}
  et~al.}{2017}]{2017MNRAS.470.4186B}
{Bah{\'e}} Y.~M.,  et~al., 2017, \mn@doi [\mnras] {10.1093/mnras/stx1403},
  \href {http://adsabs.harvard.edu/abs/2017MNRAS.470.4186B} {470, 4186}

\bibitem[\protect\citeauthoryear{{Barnes}, {Kay}, {Henson}, {McCarthy},
  {Schaye}  \& {Jenkins}}{{Barnes} et~al.}{2017a}]{2017MNRAS.465..213B}
{Barnes} D.~J.,  {Kay} S.~T.,  {Henson} M.~A.,  {McCarthy} I.~G.,  {Schaye} J.,
    {Jenkins} A.,  2017a, \mn@doi [\mnras] {10.1093/mnras/stw2722}, \href
  {http://adsabs.harvard.edu/abs/2017MNRAS.465..213B} {465, 213}

\bibitem[\protect\citeauthoryear{{Barnes} et~al.,}{{Barnes}
  et~al.}{2017b}]{2017MNRAS.471.1088B}
{Barnes} D.~J.,  et~al., 2017b, \mn@doi [\mnras] {10.1093/mnras/stx1647}, \href
  {http://adsabs.harvard.edu/abs/2017MNRAS.471.1088B} {471, 1088}

\bibitem[\protect\citeauthoryear{Benitez-Llambay}{Benitez-Llambay}{2015}]{alejandro_benitez_llambay_2015_21703}
Benitez-Llambay A.,  2015, py-sphviewer: Py-SPHViewer v1.0.0,
  \mn@doi{10.5281/zenodo.21703}, \url {http://dx.doi.org/10.5281/zenodo.21703}

\bibitem[\protect\citeauthoryear{{Crain} et~al.,}{{Crain}
  et~al.}{2015}]{2015MNRAS.450.1937C}
{Crain} R.~A.,  et~al., 2015, \mn@doi [\mnras] {10.1093/mnras/stv725}, \href
  {https://ui.adsabs.harvard.edu/abs/2015MNRAS.450.1937C} {450, 1937}

\bibitem[\protect\citeauthoryear{{Gao}, {Navarro}, {Frenk}, {Jenkins},
  {Springel}  \& {White}}{{Gao} et~al.}{2012}]{2012MNRAS.425.2169G}
{Gao} L.,  {Navarro} J.~F.,  {Frenk} C.~S.,  {Jenkins} A.,  {Springel} V.,
  {White} S.~D.~M.,  2012, \mn@doi [\mnras] {10.1111/j.1365-2966.2012.21564.x},
  \href {https://ui.adsabs.harvard.edu/abs/2012MNRAS.425.2169G} {425, 2169}

\bibitem[\protect\citeauthoryear{{Hinshaw} \& {Krauss}}{{Hinshaw} \&
  {Krauss}}{1987}]{1987ApJ...320..468H}
{Hinshaw} G.,  {Krauss} L.~M.,  1987, \mn@doi [\apj] {10.1086/165564}, \href
  {https://ui.adsabs.harvard.edu/abs/1987ApJ...320..468H} {320, 468}

\bibitem[\protect\citeauthoryear{{Hunter}}{{Hunter}}{2007}]{2007CSE.....9...90H}
{Hunter} J.~D.,  2007, \mn@doi [Computing in Science and Engineering]
  {10.1109/MCSE.2007.55}, \href
  {https://ui.adsabs.harvard.edu/abs/2007CSE.....9...90H} {9, 90}

\bibitem[\protect\citeauthoryear{{Meneghetti}, {Bartelmann}, {Dahle}  \&
  {Limousin}}{{Meneghetti} et~al.}{2013}]{2013SSRv..177...31M}
{Meneghetti} M.,  {Bartelmann} M.,  {Dahle} H.,   {Limousin} M.,  2013, \mn@doi
  [\ssr] {10.1007/s11214-013-9981-x}, \href
  {https://ui.adsabs.harvard.edu/abs/2013SSRv..177...31M} {177, 31}

\bibitem[\protect\citeauthoryear{{Meneghetti} et~al.,}{{Meneghetti}
  et~al.}{2020}]{2020Sci...369.1347M}
{Meneghetti} M.,  et~al., 2020, \mn@doi [Science] {10.1126/science.aax5164},
  \href {https://ui.adsabs.harvard.edu/abs/2020Sci...369.1347M} {369, 1347}

\bibitem[\protect\citeauthoryear{{Natarajan} et~al.,}{{Natarajan}
  et~al.}{2017}]{2017MNRAS.468.1962N}
{Natarajan} P.,  et~al., 2017, \mn@doi [\mnras] {10.1093/mnras/stw3385}, \href
  {https://ui.adsabs.harvard.edu/abs/2017MNRAS.468.1962N} {468, 1962}

\bibitem[\protect\citeauthoryear{{Perez} \& {Granger}}{{Perez} \&
  {Granger}}{2007}]{2007CSE.....9c..21P}
{Perez} F.,  {Granger} B.~E.,  2007, \mn@doi [Computing in Science and
  Engineering] {10.1109/MCSE.2007.53}, \href
  {https://ui.adsabs.harvard.edu/abs/2007CSE.....9c..21P} {9, 21}

\bibitem[\protect\citeauthoryear{{Planck Collaboration} et~al.,}{{Planck
  Collaboration} et~al.}{2014}]{2014A&A...571A..16P}
{Planck Collaboration} et~al., 2014, \mn@doi [\aap]
  {10.1051/0004-6361/201321591}, \href
  {http://adsabs.harvard.edu/abs/2014A%26A...571A..16P} {571, A16}

\bibitem[\protect\citeauthoryear{{Planelles}, {Borgani}, {Fabjan}, {Killedar},
  {Murante}, {Granato}, {Ragone-Figueroa}  \& {Dolag}}{{Planelles}
  et~al.}{2014}]{2014MNRAS.438..195P}
{Planelles} S.,  {Borgani} S.,  {Fabjan} D.,  {Killedar} M.,  {Murante} G.,
  {Granato} G.~L.,  {Ragone-Figueroa} C.,   {Dolag} K.,  2014, \mn@doi [\mnras]
  {10.1093/mnras/stt2141}, \href
  {https://ui.adsabs.harvard.edu/abs/2014MNRAS.438..195P} {438, 195}

\bibitem[\protect\citeauthoryear{{Robertson}, {Harvey}, {Massey}, {Eke},
  {McCarthy}, {Jauzac}, {Li}  \& {Schaye}}{{Robertson}
  et~al.}{2019}]{2019MNRAS.488.3646R}
{Robertson} A.,  {Harvey} D.,  {Massey} R.,  {Eke} V.,  {McCarthy} I.~G.,
  {Jauzac} M.,  {Li} B.,   {Schaye} J.,  2019, \mn@doi [\mnras]
  {10.1093/mnras/stz1815}, \href
  {https://ui.adsabs.harvard.edu/abs/2019MNRAS.488.3646R} {488, 3646}

\bibitem[\protect\citeauthoryear{{Schaye} et~al.,}{{Schaye}
  et~al.}{2015}]{2015MNRAS.446..521S}
{Schaye} J.,  et~al., 2015, \mn@doi [\mnras] {10.1093/mnras/stu2058}, \href
  {http://adsabs.harvard.edu/abs/2015MNRAS.446..521S} {446, 521}

\bibitem[\protect\citeauthoryear{{Umetsu} et~al.,}{{Umetsu}
  et~al.}{2014}]{2014ApJ...795..163U}
{Umetsu} K.,  et~al., 2014, \mn@doi [\apj] {10.1088/0004-637X/795/2/163}, \href
  {https://ui.adsabs.harvard.edu/abs/2014ApJ...795..163U} {795, 163}

\bibitem[\protect\citeauthoryear{{van der Walt}, {Colbert}  \&
  {Varoquaux}}{{van der Walt} et~al.}{2011}]{2011CSE....13b..22V}
{van der Walt} S.,  {Colbert} S.~C.,   {Varoquaux} G.,  2011, \mn@doi
  [Computing in Science and Engineering] {10.1109/MCSE.2011.37}, \href
  {https://ui.adsabs.harvard.edu/abs/2011CSE....13b..22V} {13, 22}

\bibitem[\protect\citeauthoryear{van~der Walt et~al.,}{van~der Walt
  et~al.}{2014}]{scikit-image}
van~der Walt S.,  et~al., 2014, \mn@doi [PeerJ] {10.7717/peerj.453}, 2, e453

\makeatother
\end{thebibliography}

\label{lastpage}

\end{document}